# Lightning-Induced Faults in Low-Voltage Distribution Networks via Hybrid VTS-PEEC Method

Xiaobing Xiao[1*], Xipeng Chen[1], Lei Jia[1],
Huaifei Chen[1], Lu Qu[1], Chakhung Yeung[1]

[1] State Key Laboratory HVDC Transmissions Technology, China Southern Power Grid, Guangzhou, 510700, China
544491239@qq.com

**Abstract.** As a critical component of power supply systems, low-voltage distribution networks directly affect grid stability and user power supply reliability, yet they face significant threats from lightning-induced faults. Transient simulations are more economical and adaptable for investigating lightning-induced faults in low-voltage distribution networks than experiments. A hybrid Variable Time Step (VTS)-Partial Element Equivalent Circuit (PEEC) method, has been validated in previous study, is used for Lightning-induced Electromagnetic Pulse (LEMP) simulation and fault analysis. The lightning-induced faults in extended unequal-length double-circuit low-voltage distribution networks are analyzed in this paper. The impact of lightning stroke location on overvoltage and fault risk is the primary focus of this study. Key findings indicate that, for ground strokes in front of the center of one double circuit, similar three-phase negative and bipolar oscillatory waveforms that are linked to fault initiation emerge. Closer strokes promote bipolar waveforms with the main peak negative as well as higher overvoltages and fault risk. These results provide essential insights for understanding lightning-induced fault mechanisms, thereby laying a foundation for formulating more targeted and effective lightning protection measures.

## 1    Introduction

As an important component of the low-voltage distribution network, the safe operation of the distribution network directly affects power grid stability and user power supply reliability [1,2]. Lightning strokes occur near the line induce overvoltage or overcurrent in distribution lines through electromagnetic coupling [3,4]. Compared with transmission lines, low-voltage distribution networks feature lower insulation levels, making lightning faults caused by induced lightning a significant threat. With the deep grid connection of new energy equipment, the physical complexity and lightning vulnerability of low-voltage distribution networks have become increasingly prominent [5-7]. Corresponding research on the protection against lightning faults induced by induced lightning in low-voltage distribution networks is urgently required [8,9].

Research on lightning overvoltage in low-voltage distribution networks is mainly divided into practical experiments [3,10,11] and numerical simulations [12-16]. However, due to the operational complexity of lightning experiments, numerical methods have become the primary approach to studying the propagation characteristics of direct lightning overvoltage in low-voltage distribution networks. A commonly used simulation method for low-voltage distribution networks is the numerical method combined with circuit models [17,18]. Nevertheless, circuit-based methods fail to consider the multiple coupling effects between transmission towers, lightning channels, and lines. Notably, neglecting the coupling effect between the channel and the tower will underestimate the lightning strike risk [19]. Consequently, numerical calculation methods integrated with line-field coupling models are typically adopted for calculating induced lightning faults in low-voltage distribution networks.

The full-wave numerical method, hybrid electromagnetic method, and multi-conductor transmission line method [20,21] are currently widely used numerical calculation methods for line-field coupling models. For low-voltage distribution networks, however, these methods still have limitations of varying degrees in terms of computational efficiency and modeling flexibility [22-24]. The partial element equivalent circuit method with a variable time step strategy (VTS-PEEC) provides excellent modeling flexibility for low-voltage distribution networks, and it balances computational efficiency and analysis accuracy when calculating induced lightning faults in such networks [2]. Specific details of the method, overhead line modeling, and direct lightning risk assessment using this method can be found in previous literature [2,25].

This paper adopts the VTS-PEEC method to calculate lightning fault events on low-voltage distribution lines and analyzes the lightning overvoltage waveforms along the lines, providing guidance for understanding the impact of lightning on low-voltage distribution networks and formulating more effective lightning protection measures.

This paper is structured as follows: Section 2 presents the methodology and simulation configuration. Section 3 analyzes the obtained results. Section 4 summarizes the entire paper.

## 2 Methodology and Configuration

In the VTS-PEEC hybrid model, the PEEC method is employed for model establishment. Note that, the PEEC method based on node analysis equations cannot adopt the VTS strategy. Therefore, it is necessary to convert the PEEC equations based on node analysis into state-space equations. Based on this, the state-space equation for constructing the VTS-PEEC hybrid equivalent circuit matrix is as follows:

$$x(t) = \mathbf{C}\mathbf{x}(t) + \mathbf{D}u(t) \tag{1}$$

where $\mathbf{x}(t) = \begin{bmatrix} \mathbf{V}_n(t) \\ \mathbf{I}_b(t) \end{bmatrix}$, $\mathbf{C} = \begin{bmatrix} 0 & -\mathbf{L} \\ \mathbf{P}^{-1} & 0 \end{bmatrix}^{-1} \begin{bmatrix} \mathbf{A} & \mathbf{R} \\ \mathbf{G}^{-1} & \mathbf{A}^t \end{bmatrix}$, $\mathbf{D} = \begin{bmatrix} 0 & -\mathbf{L} \\ \mathbf{P}^{-1} & 0 \end{bmatrix}^{-1}$, $\mathbf{u}(t) = \begin{bmatrix} \mathbf{V}_s(t) \\ \mathbf{I}_s(t) \end{bmatrix}$.

For the discretization of the state-space equation, the combination strategy expansion of the midpoint integration approximation method and the φ-family is adopted, with the specific expansion as follows [2,25]:

$$\dot{x}(t_1) = \sum_{i=0}^{n} \frac{1}{i!} \mathbf{C}^i (t_1 - t_0)^i x(t_0) + \mathbf{D}(t_1 - t_0) e^{(\frac{t_1-t_0}{2})\mathbf{C}} u(\frac{t_0+t_1}{2}) \quad (2)$$

The surge arrester used in this paper employs the frequency-dependent equivalent circuit model recommended by IEEE [26,27], as shown in Fig. 1, where the nonlinear voltage-current characteristics of the surge arrester are represented by two segments of nonlinear resistance.

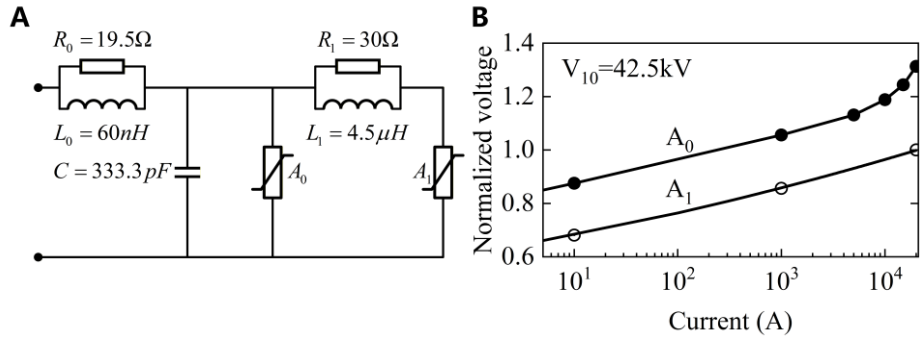

**Fig. 1.** Surge arrester model used in this paper, with (a) frequency-dependent equivalent circuit of lightning arrester, and (b) V-I characteristic curve of nonlinear resistance.

The Critical Flashover Voltage (CFO) model, recommended by IEEE [28], assesses the flashover status on an insulator, referring to the voltage level at which the insulator has a 50% probability of flashover. The threshold level is typically assumed to be 1.5 times the CFO of lines, which is defined as follows:

$$\text{CFO}_{ins} = 1.5 \times \text{CFO}_{Line} = 296 \text{ kV} \quad (3)$$

where, $\text{CFO}_{ins}$ and $\text{CFO}_{Line}$ are the CFO of insulator and line, respectively.

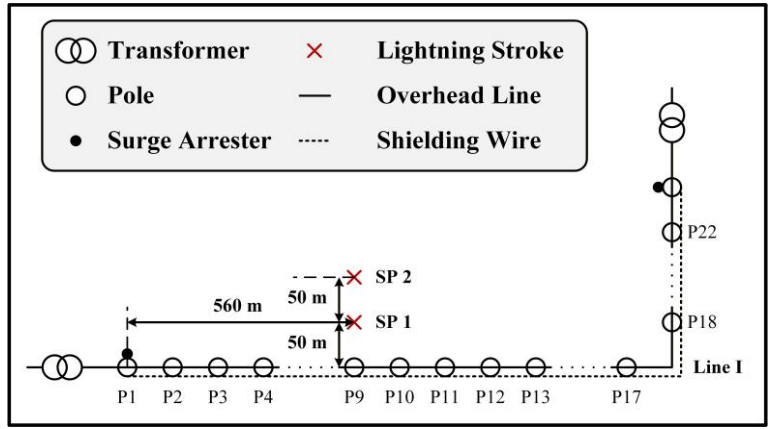

**Fig. 2.** Topology of the 10 kV distribution network for simulation.

## 3   Result and Analysis

A typical first return stroke current waveform is adopted, with the lightning stroke point 100 m away from the line, which is labeled as SP2, as shown in Fig. 2. The propagation characteristics of the induced overvoltage wave on Phase A along Circuit I are illustrated in Fig. 3. Fig. 3(a) presents the overvoltage waveforms of Pole 2 to Pole 9 on the left side of the lightning strike point, while Fig. 3(b) shows those of Pole 10 to Pole 21 on the right side. From Pole 9 to Pole 1, and from Pole 10 to Pole 21, the voltage amplitude along the line gradually decreases. The induced voltage waveform exhibits the characteristic of negative-polarity oscillation with an oscillation period of approximately 10 μs.

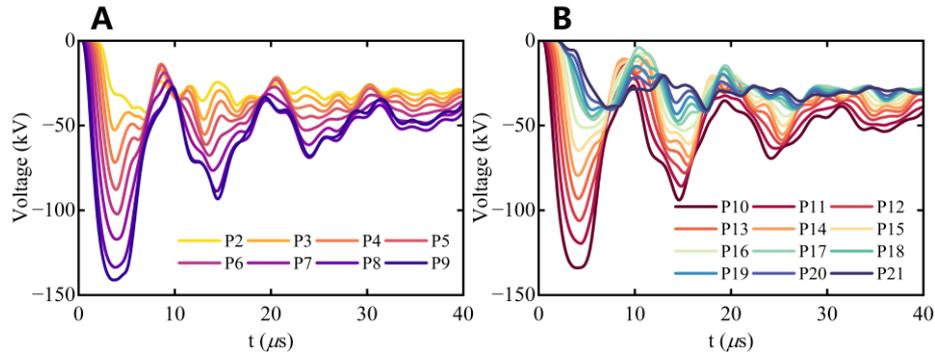

**Fig. 3.** Overvoltage propagation characteristics of Phase A along Line I with first return-stroke current waveform and lightning stroke at SP 2, where (a) and (b) are the result to the left and right of the lightning stroke point.

Assuming the wave propagates at the speed of light with a propagation distance of 3000 m, the total length of Line I is 1470 m, and the wave reflection distance is 2940 m, which is twice the total length of Line I. Such a correspondence directly indicates that the observed oscillations in the voltage waveform originate from the reflection of the lightning-induced wave at the endpoints of Line I.

Within the oscillating waveform, three distinct negative peaks can be clearly identified. These peaks are sequentially formed as the reflected wave interacts with the incident wave at different time intervals, with each peak corresponding to a specific round-trip propagation cycle of the wave along the line. Following these three negative peaks, the voltage oscillation gradually diminishes. This attenuation is attributed to the inherent resistance and leakage losses in the line, which dissipate the energy of the oscillating wave over time.

To quantify these characteristics, Table 1 systematically summarizes the values of the three negative peaks for the three-phase insulators installed on Poles 2 to 21 along the line. This tabulated data not only provides a clear reference for comparing the peak magnitudes across different poles but also lays a foundation for subsequent analyses of voltage distribution patterns and insulation withstand capability under lightning-induced conditions.

**Table 1.** Statistics on Negative Peak Amplitude of Voltage along the Line I

| Pole No. | 1st Negative peak/kV | | | 2nd Positive peak/kV | | | 3rd Positive peak/kV | | |
|---|---|---|---|---|---|---|---|---|---|
| | Phase A | Phase B | Phase A | Phase B | Phase C | Phase C | Phase A | Phase B | Phase C |
| P2 | -42.2 | -40.3 | -40.8 | -38.0 | -36.0 | -35.4 | -34.5 | -34.1 | -34.4 |
| P3 | -52.6 | -47.4 | -43.8 | -46.5 | -43.1 | -38.6 | -38.3 | -37.6 | -38.4 |
| P4 | -71.8 | -66.2 | -62.7 | -55.3 | -52.8 | -49.4 | -41.6 | -41.5 | -43.2 |
| P5 | -87.9 | -82.3 | -78.7 | -61.3 | -59.9 | -57.8 | -45.7 | -45.3 | -46.9 |
| P6 | -102.2 | -96.5 | -93.2 | -67.7 | -66.1 | -65.8 | -52.8 | -51.2 | -52.1 |
| P7 | -117.2 | -111.8 | -108.1 | -76.6 | -74.7 | -75.2 | -61.4 | -59.3 | -58.5 |
| P8 | -133.7 | -126.7 | -122.7 | -88.7 | -82.3 | -82.7 | -68.9 | -65.6 | -64.1 |
| P9 | -141.2 | -134.8 | -129.5 | -93.3 | -86.1 | -84.3 | -68.0 | -65.8 | -62.9 |
| P10 | -134.0 | -127.3 | -123.8 | -94.0 | -85.0 | -83.6 | -69.4 | -66.0 | -63.9 |
| P11 | -119.6 | -113.7 | -109.4 | -85.9 | -78.3 | -76.3 | -63.5 | -60.5 | -59.2 |
| P12 | -106.2 | -99.3 | -94.9 | -78.0 | -70.5 | -69.0 | -60.9 | -58.3 | -57.7 |
| P13 | -93.2 | -86.4 | -81.4 | -73.4 | -67.7 | -64.8 | -58.6 | -56.6 | -55.9 |
| P14 | -79.5 | -72.4 | -67.4 | -68.6 | -63.9 | -61.0 | -55.1 | -53.4 | -52.8 |
| P15 | -64.8 | -57.8 | -55.8 | -60.8 | -57.8 | -55.8 | -50.2 | -48.6 | -48.0 |
| P16 | -53.2 | -53.9 | -52.5 | -53.2 | -53.9 | -52.5 | -44.7 | -42.8 | -42.2 |
| P17 | -49.4 | -51.4 | -50.3 | -49.4 | -51.4 | -50.3 | -40.6 | -36.2 | -36.4 |
| P18 | -47.3 | -49.3 | -48.3 | -47.3 | -49.3 | -48.3 | -40.4 | -35.7 | -36.2 |
| P19 | -43.2 | -44.9 | -43.5 | -43.2 | -44.9 | -43.5 | -39.6 | -35.7 | -35.7 |
| P20 | -42.3 | -38.5 | -36.0 | -42.3 | -38.5 | -36.0 | -38.0 | -34.7 | -34.3 |
| P21 | -40.7 | -36.5 | -37.6 | -40.7 | -36.4 | -35.2 | -35.6 | -32.6 | -32.3 |

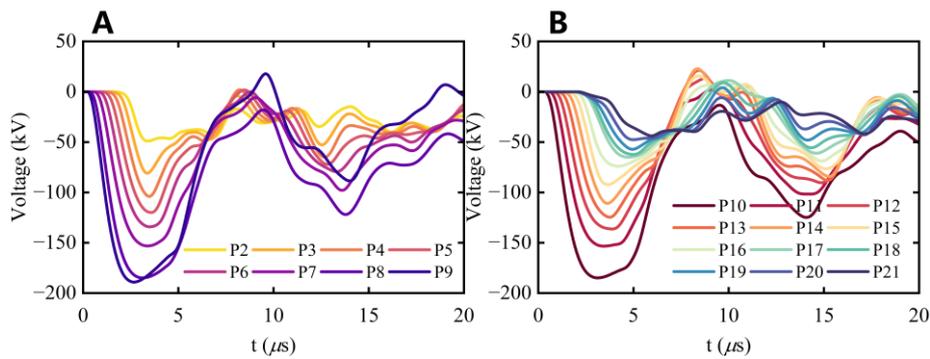

**Fig. 4.** Overvoltage propagation characteristics of Phase A along Line I with first return-stroke current waveform and lightning stroke at SP 1, where (a) and (b) are the result to the left and right of the lightning stroke point.

When a typical first return stroke current directly occurs the ground 50 m away from the line, the voltage propagation characteristics of Phase A on the towers along Line I are illustrated in Fig. 4. In contrast to the scenario in Fig. 3 where the lightning strike point is 100 m away from the line, the reduced distance intensifies the electromagnetic coupling between the lightning channel and the distribution line, prompting the voltage waveform to evolve into a distinct bipolar pattern. Despite this structural transition, the main peak remains dominated by negative polarity, which aligns with the intrinsic polarity feature of lightning-induced overvoltage in low-voltage distribution systems. Under a typical first return stroke current, the maximum negative peak amplitude of the voltage along the line is less than 200 kV, and the maximum positive peak amplitude is 22.8 kV.

## 4    Conclusion

This paper focuses on the problem that lightning-induced faults bring significant threats to low voltage distribution networks. These threats are worsened by the low insulation level of low voltage distribution networks and the increased physical complexity caused by deep grid connection of new energy equipment. Meanwhile, traditional numerical methods have limitations in computational efficiency and modeling flexibility. To solve this problem, the hybrid VTS-PEEC method, which has been validated for precision and adaptability in previous studies is applied to simulate lightning-induced faults and analyze fault mechanisms in unequal length double circuit low voltage distribution networks. The research systematically studies the impacts of lightning stroke location on overvoltage propagation and fault risk.

The findings show distinct voltage response patterns under different lightning scenarios. When lightning strokes at close range, the voltage waveform presents bipolar oscillation with a dominant negative main peak. When the lightning stroke distance increases, the voltage waveform shows unipolar negative oscillation. Both patterns are related to wave reflections at the endpoints of the line. Closer lightning strokes will increase overvoltage levels and fault potential. The quantitative results from the research include the maximum negative peak amplitude of voltage along the line, which is less than 200 kV and the distribution of voltage peaks along the line. These results not only verify the effectiveness of the VTS-PEEC method in simulating complex distribution networks but also provide important data support for evaluating the insulation withstand capacity of distribution line components and optimizing lightning protection strategies. The outcomes of this research also offer practical guidance for developing targeted lightning protection measures to improve the stability and reliability of low voltage distribution networks.

## Acknowledgement

The work leading to this paper was supported by China Southern Power Grid under project "Development of a Simulation Platform for the Transient Process of Strong Electromagnetic Pulse Coupling in New Energy Distribution Systems (Phase One) –

Subproject 1: Research on Spatio-Temporal Multi-Scale Simulation Technology for New Energy Distribution Networks Considering Probabilistic Risks of Strong Electromagnetic Pulses" (NO.GZKJXM20222352).